\documentclass[10pt]{article}
\usepackage[dvips]{graphics}
\fussy
\def\Journal#1#2#3#4{{#1}{\bf #2}, #3 (#4)}

\def\PRL{{\em Phys.~Rev.~Lett.~}}
\def\JMP{{\em J.\ Math.\ Phys.~}}
\def\CMP{{\em Commun.\ Math.\ Phys.~}}
\def\PHYS{{\em Physica\/} D}
\def\AP{{\em Ann.\ Phys.~}}
\def\PRB{{\em Phys.\ Rev.~}B}
\def\RIMS{{\em Publ.\ RIMS, Kyoto Univ.~}}

\def\ra{\rightarrow}
\def\iy{\infty}
\def\hf{{1\over 2}}

\def\be{\begin{equation}}
\def\ee{\end{equation}}

\begin{document}

\title{Universality of the Distribution Functions of Random Matrix Theory}

\author{Craig A.~Tracy\\
Department of Mathematics\\ and \\
Institute of Theoretical Dynamics\\
University of California, Davis, CA 95616, USA\\
e-mail address: tracy@itd.ucdavis.edu\\
\and
Harold Widom\\
Department of Mathematics\\ University of California, Santa Cruz,
CA 95064, USA\\
e-mail address: widom@math.ucsc.edu}

\maketitle
\begin{center}
\textit{
Dedicated to James B.~McGuire on the occasion of his sixty-fifth birthday.
}
\end{center}

\vspace{1ex}

\section{Introduction}
Statistical mechanical lattice models  are  called {\em solvable\/}
if their associated
Boltzmann weights satisfy the factorization or star-triangle equations of
McGuire~\cite{mcg}, Yang~\cite{yang} and Baxter~\cite{baxter1}.  For such
models the free energy per site and the one-point correlations in
the thermodynamic limit are expressible
in closed form~\cite{baxter2}.
There exists a
deep mathematical structure~\cite{baxter2,jm} underlying these
solvable models; and near
critical points,  a wider applicability
than one would initially expect.  This last phenomenon, called  {\em
universality\/}, has
its mathematical roots in the
strong law of large numbers and the central limit theorems of probability
theory and its physical
origins in critical phenonmena and conformal field theory.

The exact computation of $n$-point correlation functions is generally
an open problem for most ``solvable'' models. In the special case
of the  2D Ising
model, the  $n$-point  functions  (in the scaling limit)  are expressible in
terms
of solutions to integrable differential equations~\cite{wmtb,mtw,smj}.
 The Wigner-Dyson theory of random
matrices~\cite{porter,mehtaBook} is a second class of statistical models where
integrable differential equations and $n$-point correlations (or more
precisely,
level-spacing distributions) are  related.  This paper reviews some
of these relationships.

In section 2 we define the basic objects of random matrix theory (RMT).  In
section 3 we recall
the {\em bulk scaling limit\/} and the {\em edge scaling limit\/} and express
various  distribution functions in terms of Painlev\'e transcendents.  In
section 4
we discuss the universality of these distribution functions.

\section{Random Matrix Models}
In the  Gaussian models~\cite{porter, mehtaBook}, the probability
density that the eigenvalues lie in infinitesimal intervals about
the points $x_1,\ldots,x_N$ is given by
\be
P_{N\beta}(x_1,\ldots,x_N)=C_{N\beta}\; e^{-{1\over 2}\beta\sum x_i^2}\,
\prod_{j<k}|x_j-x_k|^{\beta},\label{probdensity}\ee
where  $C_{N\beta}$ is a normalization constant and
\[
\beta:=\left\{ \begin{array}{lll} 1 \mbox{\rm\quad for GOE,} \\
                               2 \mbox{\rm\quad for GUE,  }\\
                               4 \mbox{\rm\quad  for GSE.}
             \end{array}
             \right. \]
We recall that for $\beta=1$ the matrices are $N\times N$ real
symmetric, for $\beta=2$ the matrices are $N\times N$ complex
Hermitian, and for $\beta=4$ the matrices are $2N\times 2N$
self-dual Hermitian matrices. (For $\beta=4$ each eigenvalue
has multiplicity two.)  In each case the eigenvalues are real.
\par
    In RMT the probabilities
\begin{eqnarray*}
 E_{N\beta}(0;J)& := &\raisebox{-2mm}{$\displaystyle
 \int\cdots\int \atop \hspace{-2mm} x_j{\not\in} J$}
  P_{N\beta}(x_1,\ldots,x_N)\, dx_1 \cdots dx_N , \label{prob1}\\
 & = &\mbox{\rm probability no eigenvalues lie in $J$,}\nonumber
 \end{eqnarray*}
 are particularly interesting~\cite{mehtaBook}.
The simplest choices of $J$  are $(a,b)$  and
 $(t,\infty)$.
In the first instance
  the  mixed second partial derivative $E_{N\beta}(0;J)$ with respect
to $a$ and $b$ gives the spacing distribution between consecutive eigenvalues;
and in the second case,
$ F_{N\beta}(t):=E_{N\beta}(0;(t,\infty)) $
 is the
distribution function for the largest eigenvalue.

The important fact~\cite{gaudin,dyson,mehta1}
(see~\cite{tw1} for simplified proofs) is that $E_{N\beta}(0;J)$ (or its
square for $\beta=1$ or $4$) is expressible as a Fredholm
determinant.  For $\beta=2$ the kernel is a scalar kernel
acting on $J$ whereas for $\beta=1,4$, $E_{N\beta}(0;J)^2$
equals a Fredholm determinant of $2\times 2$ matrix kernel acting on $J$.

\section{Painlev\'e Representations}
The representation of $E_{N\beta}(0;J)$, or its square, as a
Fredholm determinant of an operator with kernel
of a special form  permits the determinant to
be expressed in terms of Painlev\'e functions~\cite{jmms,airy,fred,orthogonal}.

\subsection{Bulk Scaling Limit}
Denote the density of eigenvalues at the point $x_0$ by $\rho(x_0)$.   It
is customary in the limit $N\rightarrow\infty$ to scale distances so that
 the resulting density is one.  Precisely, we define $\xi=\rho_N(x_0)(x-x_0)$,
$x_0$ independent of $N$,
and consider the limit $N\rightarrow\infty$, $x\rightarrow x_0$, such that
$\xi$ is fixed.  By requiring
$x_0$ to be independent of $N$, we are choosing a point in the ``bulk'' of the
spectrum and
are examining the local statistics of the eigenvalues in some small
neighborhood
of the point $x_0$.  In this limit,
and for $\beta=2$,  we are  led to the Fredholm
determinant of the operator on $L^2(0,s)$ whose kernel is the famous
sine-kernel~\cite{gaudin,mehtaBook}
\[
K(\xi,\xi^\prime):={1\over \pi} {\sin\pi(\xi-\xi^\prime)\over \xi-\xi^\prime}.
\]
Observe that the kernel is translationally invariant and independent of the
point $x_0$.

It is a result of Jimbo et al.~\cite{jmms} that
\[
\det\left(I-\lambda K\right)=\exp\left(\int_0^{\pi s} {\sigma(x;\lambda)\over
x}\, dx\right),
\]
where $\sigma$ satisfies the differential equation
\[
\left(x\sigma^{\prime\prime}\right)^2+4\left(x\sigma^\prime-\sigma\right)\left(
x\sigma^\prime-\sigma+(\sigma^\prime)^2\right)=0
\]
with the
 boundary condition
 $ \sigma(x;\lambda)\sim -{\lambda\over \pi}\, x$ as $x\rightarrow 0$.
(Here $K$ is the operator whose kernel is the sine-kernel acting on
$L^2(0,s)$.)
The differential equation is the
``$\sigma$ representation'' of the $P_V$ equation~\cite{jmms,jm2}.
For other   proofs of this result see~\cite{mehta2,dyson2,intro}.

If  $E_\beta(0;s)$ denotes the limiting value of $E_{N\beta}(0;(-t,t))$ in the
bulk
scaling limit with the scaled length of $J$ set equal to $s$,
then~\cite{mehtaBook,intro,orthogonal}
\begin{eqnarray*}
E_1(0;s)&=&\det\left(I-K_+\right), \\
E_2(0;s)&=&\det\left(I-K\right), \\
E_4(0;s/2)&=&{1\over 2}\left(\det\left(I-K_+\right) +
\det\left(I-K_-\right)\right)
\end{eqnarray*}
where $K_\pm$ are the operators with kernels $K(x,y)\pm K(-x,y)$.
The $\det(I-K_\pm)$ can be expressed in terms of $\det(I-K)$~\cite{jmms,intro}:
\[
\det\left(I-K_\pm\right)^2=\det\left(I-K\right)\exp\left(\mp\int_0^s\sqrt{
-{d^2\over dx^2}\log\det(I-K)}\,dx\right).
\]
These formulas for $E_\beta(0;s)$ are well adapted for producing graphs
of $E_\beta(0;s)$ and the level-spacing densities
$p_{\beta}(s)=d^2E_{\beta}(0;s)/ds^2$ once one
knows  the large $s$ asymptotics of $\sigma(s;1)$
and $E_\beta(0;s)$~\cite{btw,intro,kitaev}.

\subsection{Edge Scaling Limit}
\par
 The  limiting law, called the Wigner semi-circle law, is well known
 \[ \lim_{N_\beta\ra\iy} {1\over 2 \sigma \sqrt{N_\beta}}
\rho_N\left(2\sigma\sqrt{N_\beta}\, x\right) =
 \left\{ \begin{array}{ll} {1\over \pi}\sqrt{1-x^2} & \vert x \vert < 1, \\
 				0 & \vert x \vert > 1.
 				\end{array}\right. \]
 Here $\sigma$, $\sigma/\sqrt{2}$, $\sigma/\sqrt{2}$ (for $\beta=1,2,4$,
respectively)
 is the standard deviation of the Gaussian distribution in the off-diagonal
 elements and
\[ N_\beta=\left\{ \begin{array}{lr} N,\> \beta=1, \\
                               N,\> \beta=2,\\
                               2N+1,\>\beta=4.
             \end{array}
             \right. \]
   For the normalization here, $\sigma=1/\sqrt{2}$.
 Less known is that
 the distribution function of the largest eigenvalue satisfies~\cite{BY}:

 \[ F_{N\beta}\left(2\sigma\sqrt{N_\beta}+x\right) \ra \left\{\begin{array}{ll}
0 &
\mbox{if $x<0$,} \\
                                             1 & \mbox{if $x>0$,}
                                             \end{array}\right. \]
  as $N\ra\iy$.
  The edge scaling variable, $s$,  defined by
  \[ t= 2\sigma\sqrt{N_\beta} + {\sigma\> s \over N_\beta^{1/6} },\]
gives the scale of the fluctuations at the edge of the
spectrum~\cite{brezin,forrester,airy}.
 As  $N\ra\iy$ with  $s$ fixed~\cite{airy,fred}
 \be F_{N2}(t)\ra \exp\left(-\int_s^\iy (x-s) q(x)^2 \, dx\right) =:
F_2(s)\label{F2}\ee
  where $q$ is the solution to the $P_{II}$ equation
\[ q''=s\,q+2\,q^3\]
 satisfying the condition
 \[ q(s)\sim {\rm Ai}(s) \ \ \mbox{as}\ \ s\ra\iy, \]
 with $\mbox{Ai}$ the Airy function.
 \par
For $\beta=1,4$ the results are~\cite{orthogonal}
\[ F_{N1}(t)^2\ra
F_1(s)^2=F_2(s)\,\exp\left(-\int_s^{\iy}q(x)\,dx\right)\]
and
\[ F_{N4}(t)^2\ra
 F_4(s/\sqrt{2})^2=F_2(s)\,\cosh^2\left(\hf\int_s^{\iy}q(x)\,dx\right).\]
 Table 1 gives some statistics of $F_\beta$ and
 Figure 1 shows the densities $f_\beta(s)=dF_\beta/ds$.

\begin{table}
\begin{center}
\caption{The mean ($\mu_\beta$),  standard deviation ($\sigma_\beta$),
skewness ($S_\beta$) and  kurtosis ($K_\beta$) of $F_\beta$.}
\vspace{2ex}
\begin{tabular}{|l|cccc|}\hline
$\beta$ & $\mu_\beta$ & $\sigma_\beta$ & $S_\beta$ & $K_\beta$ \\  \hline
1 & -1.20653 & 1.2680 & 0.293 & 0.165 \\
2 & -1.77109 & 0.9018 & 0.224 & 0.093 \\
4 & -2.30688 & 0.7195 & 0.166 & 0.050 \\ \hline
\end{tabular}
\end{center}
\end{table}
\begin{figure}
\vspace{-1in}
\begin{center}
\caption{The probability density $f_\beta$
 of the largest eigenvalue, $\beta=1,2,4$.}
\vspace{2ex}
\resizebox{7cm}{6cm}{\includegraphics{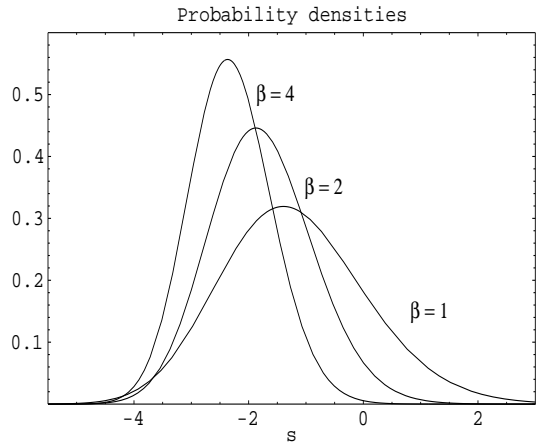}}
\end{center}
\end{figure}

\section{Universality}
The universality of the distribution functions of RMT largely
accounts for its success in
 a wide range of applications, see, e.g.~\cite{guhr}.  However,
 this universality has only been \textit{proved} in a small number of cases.
 The evidence for universality is largely numerical.

 There are two types of universality results.
  The first, and the
easier
 to establish rigorously, consists in modifying the random matrix
 model itself by, say, replacing the Gaussian $x^2/2$ appearing
 in (\ref{probdensity}) by  an ``arbitrary'' potential
 $V(x)$.  Changing the potential $V$ changes the density of
 eigenvalues $\rho$, but several authors have established that
 the bulk scaling limit results in the sine-kernel and at soft edges
 one generically obtains the Airy universality class.  (Fine tuning at the edge
 can result in different universality classes, see, e.g.~\cite{brezin}.)

 The second type of universality, and the one first envisioned by
 Wigner, asserts~\cite{bohigas}
  that for a classical, ``fully'' chaotic Hamiltonian
 the corresponding quantum system has a level spacing distribution
 equal to $p_\beta(s)$ in the bulk.
  (The symmetry class determines which ensemble.)
 A nice numerical example of this \textit{quantum chaos} is
Robnik's work~\cite{robnik} on  chaotic billiards.

 Here we discuss three other examples of RMT universality.

 \subsection{Zeros of the Riemann Zeta Function}
 Work by Montgomery~\cite{montgomery} followed
 by extensive numerical calculations by Odlyzko~\cite{odlyzko} on zeros
 of the Riemann zeta function have given convincing numerical evidence
 that the normalized consecutive spacings follow the GUE distribution, see
Figure 2 (the
 GUE Hypothesis).
  Rudnick and Sarnak~\cite{rudnick} have proved
a restricted form of this hypothesis.
\begin{figure}
\vspace{-2cm}
\begin{center}
\caption{Data for nearest neighbor spacings among 1,041,600 zeros
of the Riemann zeta function near the $2\times 10^{20}$-th zero
are plotted together with the GUE spacing density.  Courtesy of
Andrew Odlyzko~\cite{odlyzko}.}
\vspace{-3ex}
\resizebox{9cm}{9cm}{\includegraphics{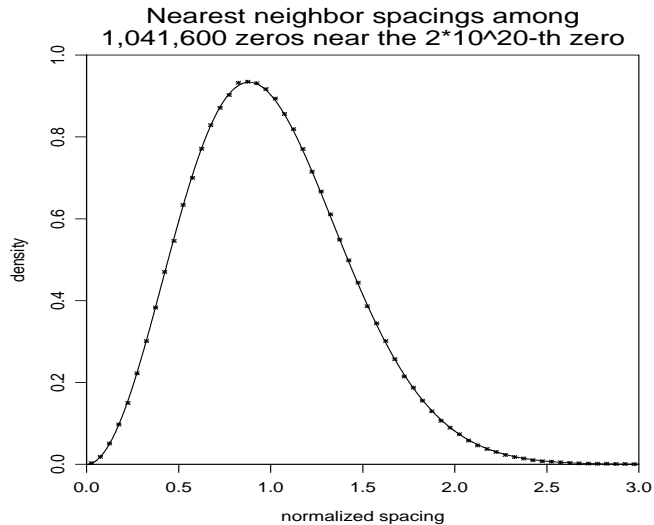}}
\end{center}
\vspace{-1cm}
\end{figure}

 \subsection{Eigenvalues of Adjacency Matrices of Quasiperiodic Tilings}
 The discovery of quasicrystals has made the study of statistical
 mechanical
 models whose underlying lattice is quasiperiodic of considerable
 interest to physicists.  In particular,  in order to understand
 transport properties, tight binding models have been defined
 on various quasiperiodic lattices.  One such study
 by Zhong et al.~\cite{grimm}  defined a simplified tight binding model
 for the octagonal tiling of Ammann and Beenker.  This quasiperiodic
 tiling consists of squares and rhombi with all
 edges of  equal lengths (see, e.g., \cite{grimm})
  and has a $D_8$ symmetry around the central vertex.
 On this tiling the authors take as their Hamiltonian the adjacency
 matrix for the graph with free boundary conditions.  The largest
 lattice they consider has 157,369 vertices.  The matrix splits
 into ten blocks according to the irreducible representations
 of the dihedral group $D_8$.  For each of these ten independent subspectra,
 they compare the empirical cumulative
 distribution of the normalized   spacings of the
 consecutive eigenvalues with the
 integrated density
 $ I_1(s)=\int_s^\infty p_1(x)\, dx $
 where $p_1$ is the GOE level spacing density.  In Figure 3 we have
 reproduced a small portion of their data for one
 such subspectrum together with  $I_1$.
  \begin{figure}
\vspace{-2cm}
\begin{center}
\caption{Data for nearest neighbor,  cumulative,
normalized spacings of eigenvalues
of the adjacency matrix for a quasiperiodic octagonal tiling
 are plotted together with the GOE integrated spacing  density $I_1(s)$.
Data are  from one independent subspectrum of a $D_8$-symmetric
octagonal patch of a tiling with 157,369 vertices. Courtesy of
Zhong et al.~\cite{grimm}.}
\vspace{2ex}
\resizebox{7cm}{6cm}{\includegraphics{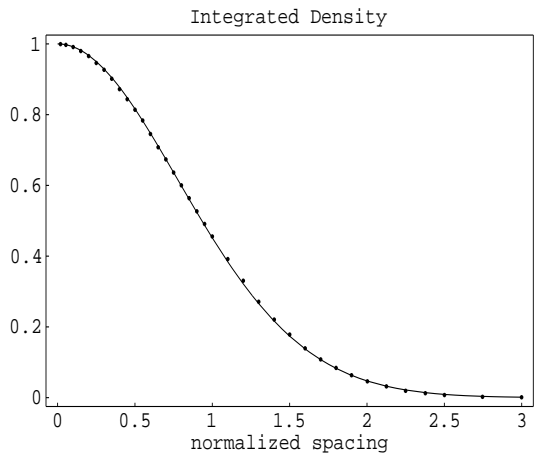}}
\end{center}
\vspace{-5mm}
\end{figure}

 \subsection{Distribution of the Length of the Longest Increasing Subsequence
of
 Random Permutations}
 An old problem going back to Ulam asks for the limiting behavior
 of the length of the longest increasing subsequence of a random permutation.
 Precisely, if $\pi$ is a permutation of $\{1,2,\ldots,N\}$, we say that
 $\pi(i_1), \cdots,\pi(i_k)$ is an increasing subsequence in $\pi$ if
 $i_1<\cdots<i_k$
 and $\pi(i_1)< \cdots<\pi(i_k)$. Let $\ell_N(\pi)$
 be the length of the longest
 increasing subsequence of $\pi$.  We take each permutation to be equally
 likely thus making $\ell_N$ a random variable.  The problem then is to
 understand the distribution
 of $\ell_N$.  This problem has a long history (see, e.g.~\cite{baik}) but its
 connection with RMT is recent~\cite{odlyzko2,rains}.  In a recent paper, Baik,
Deift
 and Johansson~\cite{baik}
 have proved the following remarkable result:
 \vspace{2ex}

 \noindent\textbf{Theorem:}
 Let $S_N$ be the group of all permutations of $N$ numbers with uniform
distribution
 and let $\ell_N(\pi)$ be the length of the longest increasing subsequence of
$\pi\in S_N$.
 Let $\chi$ be a random variable whose distribution function is $F_2$,
 the distribution function for the largest eigenvalue in the GUE in the edge
scaling
 limit~(\ref{F2}).  Then, as $N\ra\infty$,
 \[ \chi_N:={\ell_N - 2\sqrt{N}\over N^{1/6}}\ra \chi \]
 in distribution, i.e.
 \[ \lim_{N\ra\infty} \textrm{Prob}\left(\chi_N:=
 {\ell_N - 2\sqrt{N}\over N^{1/6}}\leq s\right)=F_2(s)\]
 for all $s\in {\bf R}$.

\section*{Acknowledgments}
The authors wish to thank Andrew Odlyzko and Uwe Grimm for providing them with
their data and allowing us to reproduce it here. We thank J.~Baik for sending
us~\cite{baik} prior to publication.
The first author
acknowledges helpful conversations with Bruno Nachtergaele
and Marko Robnik.
This work was supported in part by the National Science
        Foundation through grants DMS--9802122 and
        DMS--9732687, and this support is gratefully acknowledged.

\end{document}